\documentclass[aps,pra,preprintnumbers,nofootinbib,onecolumn,a4paper]{revtex4}
\usepackage{bm}
\usepackage{latexsym}
\usepackage{dcolumn}
\usepackage{amsmath,amsfonts,amssymb}
\usepackage{graphicx,epsfig}
\usepackage{amsthm}

\def\be {\begin{equation}}
\def\ee {\end{equation}}
\def\bea {\begin{eqnarray}}
\def\eea {\end{eqnarray}}
\def\bc {\begin{center}}
\def\ec {\end{center}}
\def\bfg {\begin{figure}}
\def\efg {\end{figure}}
\def\bi {\begin{itemize}}
\def\ei {\end{itemize}}

\def\la {\label}
\def\le {\left}
\def\ri {\right}

\def\beq{\begin{equation}}
\def\eeq{\end{equation}}
\def\br{\begin{eqnarray}}
\def\er{\end{eqnarray}}
\newcommand{\eel}[1] {\label{#1}\end{equation}}

\newcommand{\bdm}{\begin{displaymath}}
\newcommand{\edm}{\end{displaymath}}

\begin{document}
\title{Relativistic particle in a box: Klein-Gordon vs Dirac Equations}

\author{Pedro Alberto $^1$}\email[Email: ]{pedro.alberto@uc.pt}
\author{Saurya Das $^2$} \email[Email: ]{saurya.das@uleth.ca}
\author{Elias C. Vagenas $^3$} \email[Email: ]{elias.vagenas@ku.edu.kw}
\affiliation{$^1$Physics Department and CFisUC, University of Coimbra, P-3004-516
Coimbra, Portugal}

\affiliation{$^2$Theoretical Physics Group and Quantum Alberta, Dept of Physics and
Astronomy, University of Lethbridge, 4401 University Drive,
Lethbridge, Alberta, Canada T1K 3M4}

\affiliation{$^3$~Theoretical Physics Group, Department of Physics, Kuwait University,
P.O. Box 5969, Safat 13060, Kuwait}
\begin{abstract}
\noindent
The problem of a particle in a box is probably the simplest problem in quantum mechanics which allows for significant
insight into the nature of quantum systems and thus is a cornerstone in the teaching of quantum mechanics.
In relativistic quantum mechanics this problem allows also to highlight the implications of special relativity for
quantum physics, namely the effect that spin has on the quantized energy spectra.
To illustrate this point, we solve the problem of a spin zero relativistic particle in a one- and three-dimensional box
using the Klein-Gordon equation in the Feshbach-Villars formalism. We compare the solutions and
the energy spectra obtained with the corresponding ones from the Dirac equation
for a spin one-half relativistic particle.  We note the similarities and differences, in particular
the spin effects in the relativistic energy spectrum.
As expected, the non-relativistic limit is the same for both kinds of particles, since, for a particle in a box,
the spin contribution to the energy is a relativistic effect.
\end{abstract}

\maketitle

%
%

\section{Introduction}

The problem of a particle confined in a one-dimensional infinite square well
potential lies at the heart of non-relativistic quantum mechanics, being the simplest problem that
illustrates how the wave nature of bound particles implies that their energy is
quantized. So it is no surprise that introductory quantum mechanics courses present and discuss it as
a powerful pedagogical tool to introduce the students the particular features of the spectra of
quantum systems.

The generalization of this problem to three dimensions is used for the statistical description
of a gas of quantum particles (see, for example, ref. \cite{Pathria}) and thus is presented also in
statistical physics courses.

In relativistic quantum mechanics, where quantum mechanics and special relativity are combined, new features of the energy spectra
appear coming from the relativistic energy-momenta relations and spin, as well as the difficulties which arise when one tries to describe one-particle quantum systems similar to ones used in non-relativistic
quantum physics courses. The problem of a relativistic particle in a box is again a direct way to address those questions and thus is an useful
tool to discuss, in advanced quantum mechanics courses, the issues arising when one extends quantum mechanics to incorporate special relativity
(see, for instance, \cite{Li_Blinder}).

The relativistic formulation and solution of the problem of a
spin-1/2 fermion with mass $m$ confined in a one-dimensional
square well potential was done by Alberto, Fiolhais and Gil
\cite{afg} and then extended to a three-dimensional square well potential by Alberto, Das and
Vagenas \cite{adv}.

The problem of a relativistic
particle confined in an infinite square well potential is traditionally not
considered in the textbooks of Relativistic Quantum Mechanics, even in the
most comprehensive ones such as Greiner's \cite{Greiner}, which considers only the case of a finite square
well with a vector (energy coupled) potential. However, as discussed in \cite{afg,adv}, in relativistic
problems one cannot simply let the well height go to infinity, because this would lead to problems
related to the Klein paradox, i.e., the emergence of negative energy solutions and thus compromise
the one-particle picture that one wishes to retain. We address this problem again below.
Other problems particular to the relativistic problem concern the boundary conditions, which cannot be
the same as in the non-relativistic problems, as discussed in refs. \cite{VAlonso} and
\cite{menon}.

Here we extend further the problem of relativistic particle in a box to
spin-0 particles, i.e., solutions of the Klein-Gordon equation in an infinite well,
both in one- and three-dimensional space.
To this end, we use the Feshbach-Villars formalism \cite{fv}, by which one can have
a Schr\"odinger-type equation of motion, linear in the time derivative and thus having
a Hamiltonian defined for relativistic spin-0 particles.
This allows us to follow a similar procedure to obtain an infinite well solution
as was done for the Dirac equation, in particular achieving a clear separation between particle and anti-particle
solutions. Although not common, the Feshbach-Villars formalism has been applied in other contexts,
as in computing creation rates for particle-antiparticle pairs produced
by a supercritical force fields \cite{Cheng_pra}.
\par\noindent
Finally, we compare the solutions obtained with those for the Dirac equation and draw the conclusions.

%
\section{Klein-Gordon equation in Feshbach-Villars formalism}
%
%
The Feshbach-Villars formalism (FVF) consists essentially in obtaining a first-order differential equation in time from
the original second-order Klein-Gordon equation.
This is achieved by defining two wave functions $\varphi$ and $\chi$.
Following the original Feshbach-Villars paper \cite{fv}
but with a slight different notation and definitions, and also inspired by the textbook of Greiner \cite{Greiner}, one
may write
\bea
\psi &=& \varphi + \chi  \la{psia} \\
i\hbar\frac{\partial\psi}{\partial t}&=&mc^2(\varphi - \chi) \la{psib}
%
\eea
where $\psi$ is the conventional Klein-Gordon wavefunction.
If now one defines the two-component spinor
%
\beq
\Psi = \left( \begin{array}{c}
\varphi \\[2mm]
\chi
 \end{array} \right)\ ,
\eeq
the free Klein-Gordon equation can be written as a Schr\"odinger-type equation of motion
\bea
 i \hbar \frac{\partial \Psi}{\partial t}&=&H\Psi \la{kgeq1}\\
H &=& \le( \tau_3 + i\tau_2 \ri) \frac{\hat{\vec p}^{\;2}}{2m} + mc^2 \tau_3 ~, \la{kgham1}
\eea
where $\tau_k$ ($k=1,2,3$) are Pauli matrices and $\hat{\vec p}$ is the momentum operator,
given as $\hat{\vec p}=-i\hbar\nabla$. 
Note that the Hamiltonian (\ref{kgham1}) is similar to the one of a non-relativistic particle
plus its rest energy.
Equation (\ref{kgeq1}) is equivalent to two coupled equations for $\varphi$ and $\chi$
\bea
 i \hbar \frac{\partial \varphi}{\partial t} &=&\frac{\hat{\vec p}^{\;2}}{2m}(\varphi + \chi)+mc^2\varphi   \\
 i \hbar \frac{\partial \chi}{\partial t}&=&-\frac{\hat{\vec p}^{\;2}}{2m}(\varphi + \chi)-mc^2\chi \ .
\eea
One can check, from these last equations and equations (\ref{psia}) and (\ref{psib}), that one gets the free Klein-Gordon equation for $\psi$
\beq
\frac1{c^2}\frac{\partial^2 \psi}{\partial t^2}-\nabla^2\psi+\frac{m^2c^2}{\hbar^2}\psi=0 \ .
\eeq
\par\noindent
The conserved charge and current assume the following forms
\bea
\rho &=&\varphi^\star \varphi - \chi^\star\chi =\Psi^\dagger \tau_3 \Psi \\
\notag
\vec J &=&
\frac{\hbar}{2im}\le[
(\varphi^* + \chi^*)\nabla(\varphi + \chi)
- (\varphi + \chi)\nabla(\varphi^* + \chi^*)\ri]\\
&=& \frac{\hbar}{2im}\le[
\Psi^\dagger\tau_3 \le( \tau_3 + i\tau_2 \ri) \nabla\Psi
- \le( \nabla \Psi^\dagger \ri) \tau_3 \le( \tau_3 + i \tau_2 \ri) \Psi
\ri]
\ .\la{kgcurrent}
\eea
\par\noindent
Similarly to spin-1/2 particles, there is a charge conjugated spinor given by
\beq
\Psi_c=\left( \begin{array}{c}
\chi^* \\[2mm]
\varphi^*
 \end{array} \right)=\tau_1\Psi^*
\eeq
which changes particle into anti-particle solutions and vice-versa. One can check
that the charge conjugated density and current are
\bea
\rho_c &=&-\rho  \\
\vec J_c &=&- \vec J\ ,\la{kgcurrent_c}
\eea
meaning that these should be interpreted as a charge density and charge current, as is well-known
to be the case for the Klein-Gordon equation. One may also note that here $\tau_3$ is related to the particle charge, so
that the physical significance of Pauli matrices $\tau_i$  is similar to that of the Pauli matrices used, for instance,  
for the isospin degree of freedom of nucleons.
%
%
%
\section{Solution of the Klein-Gordon Equation for squared well potentials}
%
%
The solution of the free Klein-Gordon equation (\ref{kgeq1}) in the FVF which has good linear momentum, i.e.,
which is an eigenstate of the linear momentum operator, is given by
\cite{fv}
\beq
\Psi^{\pm}_{\vec p}(\vec x,t )=A_{\pm} \left( \begin{array}{c}
\varphi_0^{\pm} (\vec p ) \\[2mm]
\chi_0^{\pm} (\vec p)
 \end{array} \right) e^{ i(\mp E_p t-\vec p \cdot \vec x)/\hbar} \equiv
  A_{\pm}\Psi_0^{\pm}(\vec p)e^{ i(\mp E_p t-\vec p \cdot \vec x)/\hbar}
\label{psi_p}
\eeq
where $\pm$ denotes the positive (negative) energy solutions, i.e.,
$E=\pm E_p$, with $E_p = \sqrt{p^2c^2 + m^2 c^4}$, $A_{\pm}$ are normalization constants, and functions
$\varphi_0^{\pm} (\vec p )$ and $\chi_0^{\pm} (\vec p )$ are written as
%
\bea
\label{phi0}
\varphi_0^{\pm} (\vec p) &=& \frac{\pm E_p+mc^2}{2\sqrt{mc^2 E_p}} \\
\label{chi0}
\chi_0^{\pm} (\vec p) &=& \frac{mc^2\mp E_p}{2\sqrt{mc^2 E_p}}~.
\eea
Note that $\varphi_0^\pm(\vec p )$ and $\chi_0^\pm(\vec p )$ depend on $\vec p$ only through its magnitude
$|\vec p|$ and $\big[\varphi_0^{\pm}(\vec p )\big]^2-\big[\chi_0^{\pm}(\vec p )\big]^2 =\pm 1$. We also have
$\Psi^{\pm}_{\vec p}(\vec x,t )_c=\Psi^{\mp}_{-\vec p}(\vec x,t )$.
%
%
\subsection{Solution for one-dimensional squared well potential}
%
%
In order to obtain the solution of the Klein-Gordon equation for an infinite squared well potential, one must tackle the
problem that arises with the one-particle description in Relativistic Quantum Mechanics when particles are subject to
very strong external potentials. This shows up clearly when one computes the transmission of a plane wave through a
vector (energy-coupling) potential barrier, known as the Klein paradox. While it takes a different form for spin-1/2 particles than for spin-0 particles, one common feature is that in both cases a
strong vector potential leads to a non-zero transmission to a classically forbidden region, which is related to particle/anti-particle pair
production. A good historical review can be found in
ref. \cite{cd_K_P}. In order to avoid this problem and retain the one-particle quantum description of spin-0 particles with wave functions which are
solutions of the Klein-Gordon equation, one must consider the one-dimensional potential defined in a one-dimensional box
${\mathbb L} \equiv \{ 0 \leq x \leq L \}$ with length $L$ as a potential invariant under Lorentz transformations, i.e.,
a position dependent mass term, as follows
\beq
m(x)=\left\{
\begin{array}{lc}
  m &,\  x \in {\mathbb L}\\[2mm]
  M \to \infty &,\  x \notin {\mathbb L}
\end{array}\right.\ ,
\label{1d_pot}
\eeq
as was done in the Dirac equation case \cite{afg}. In this case, one can consider the time-independent solutions of the Klein-Gordon equation in the FVF, that is, one has
\beq
H\tilde\Psi(\vec x)=E\tilde\Psi(\vec x) \quad {\rm where}\quad \Psi(\vec x,t)=e^{-i E t/\hbar}\tilde\Psi(\vec x) \  .
\eeq
\par\noindent
From equation (\ref{psi_p}), the general solution for a time-independent positive-energy solution in this one-dimensional box is given by,
with $p>0$,
\beq
\Psi_B(x)=\left\{
\begin{array}{cc}
 A \Psi_0^{+}(p)e^{ ip x/\hbar}+B\Psi_0^{+}(p)e^{- i p x/\hbar} &,\  x \in {\mathbb L}\\[2mm]
0 &\ ,  x \notin {\mathbb L}
\end{array}\right.\ .
\label{1d_wf}
\eeq
The continuity of the wave function at the box borders yields the box boundary conditions
$\Psi_B (0) = 0 = \Psi_B(L)$. Together with the normalization
$\int_{\mathbb L} \rho \,dx =1$, one has for the wave function inside the box
\beq
\Psi_B (x)= \sqrt{\frac{2}{L}}\Psi_0^+(p) \sin(p\,x/\hbar) \quad {\rm with}\quad p\,L=n\pi\hbar\quad n=1,\ldots \ .
\eeq
This gives rise to the quantized energy
\bea
E_B(n) &=&\sqrt{\frac{n^2\pi^2\hbar^2 c^2}{L^2} + m^2c^4 } \\
&\approx&
mc^2 + \frac{n^2 \pi^2 \hbar^2}{2m L^2} + \dots 
\eea
where a small $p/(mc)=n\pi\hbar/(L mc)$ expansion has been made in the last line, with the second term being that
which arises by solving the Schr\"odinger equation, $mc^2$ is the rest energy,
and $\dots$ represent second and higher order terms.

The usual Klein-Gordon wavefunction $\psi$ and its time derivative, given by (\ref{psia})
and (\ref{psib}) respectively, can be recovered noting that in those equations
\bea
\varphi&=&\sqrt{\frac{2}{L}}\,\varphi_0^+(p_n)\, e^{-i E_B(n) t/\hbar}\sin(p_n x/\hbar)\\
\chi&=&\sqrt{\frac{2}{L}}\,\chi_0^+(p_n)\, e^{-i E_B(n) t/\hbar}\sin(p_n x/\hbar)
\eea
in which $p_n=n\pi\hbar/L$ and $\varphi_0^+(p_n)$ and $\chi_0^+(p_n)$ are given
respectively  by
(\ref{phi0}) and (\ref{chi0}) with $E_p$ replaced by $E_B(n)$.

Finally, the current as given by (\ref{kgcurrent}) vanishes in this case, since it is composed
of a left and a right moving wave with symmetric amplitudes,
satisfying at the same time current conservation throughout all space.
%
%
\subsection{Solution for the three-dimensional squared well potential}
%
%
For a three dimensional box defined by ${\mathbb V} \equiv \{ 0 \leq x_i \leq L_i \},~i=1,2,3$,
where $L_i$ are the lengths of the box in each dimension,
the scalar infinite well potential would be given by
\begin{equation}
m(\vec x)=\left\{
\begin{array}{cc}
  m &,\  \vec x\in {\mathbb V}\\[2mm]
  M \to \infty &,\   \vec x\not\in {\mathbb V}
\end{array}\right.\ .
\label{3d_pot}
\end{equation}
Since the potential inside the box is constant, we expect that the wave function in the box would be a linear combination of
wave functions with good momenta of the type (\ref{psi_p}). One may note also that the potential (\ref{3d_pot}) can be written
as separable potential in terms of the one-dimensional well potentials: $m(\vec x)=m+(M-m)\prod_{i=1}^3[\theta(x_i-L_i)+\theta(-x_i)]$, where $\theta(x)$ is the Heaviside step function.
This means that the space-dependent part of the wave function may be written as a product of a linear combination of
one-dimensional plane waves travelling both ways in each direction, or a linear combination of the eight plane waves
\beq
e^{i \sum_{i=1}^3 \epsilon_i p_i x_i/\hbar}\quad {\rm where}\quad \epsilon_i=\pm 1\ .
\eeq
Since the magnitude of the momenta of all of these plane waves $(\pm p_1, \pm p_2, \pm p_3)$
is the same, one can write the general time-independent positive-energy wave function as
\beq
\Psi_B(\vec x)=\left\{
\begin{array}{cc}
 \displaystyle
 \Psi_0^{+}(\vec p)\sum_{\vec\epsilon}A_{\vec\epsilon}\,
 e^{i \sum_{i=1}^3 \epsilon_i p_i x_i/\hbar} &,\   \vec x \in {\mathbb V}\\[2mm]
0 &,\   \vec x \notin {\mathbb V}
\end{array}\right.\ ,
\label{3d_wf}
\eeq
where $\vec\epsilon$ represents all the 8 sets of values of $(\epsilon_1,\epsilon_2,\epsilon_3)$,
that is, $(\pm 1,\pm 1,\pm 1)$ and $A_{\vec\epsilon}$ is
the amplitude for each plane wave.
Note the formal similarity with the Dirac wave function for a relativistic spin-$1/2$
particle in a three dimensional box, equation (33) of \cite{dva} and equation (24) of \cite{adv}.

The boundary conditions and normalization are now
\bea
\Psi_B (0,0,0) &=& 0 = \Psi_B (L_1,x_2,x_3) = \Psi_B (x_1,L_2,x_3) = \Psi_B (x_1,x_2,L_3) \\
\int_{\mathbb V} \rho \,d^3\vec x &=& 1
\eea
giving rise to the interior wave function
\beq
\Psi_B(\vec x)=\sqrt{\frac{2^3}{V}}~\Psi_0^+ (\vec p)\prod_{i=1}^3 \sin(p_i x_i/\hbar)\ ,
\eeq
where $V=L_1 L_2 L_3$ is the volume of the box and one has $p_i L_i/\hbar = n_i \pi$, with $n_i=1,\ldots$, and $i=1,2,3$.

\bea
\label{KG_box3D}
E_B(\vec n) &=& \sqrt{\sum_{i=1}^3 \frac{ n_i^2\pi^2\hbar^2 c^2}{L_i^2} + m^2 c^4 } \\
&\approx& mc^2 + \sum_{i=1}^3 \frac{n_i^2 \pi^2\hbar^2}{2mL_i^2}+ \dots
\eea
where $\vec n=(n_1,n_2,n_3)$, and the wave functions $\varphi$ and $\chi$ are now
\bea
\varphi&=&\sqrt{\frac{2^3}{V}}\,\varphi_0^+(\vec p(\vec n))\, e^{-i E_B(\vec n) t/\hbar}
\prod_{i=1}^3 \sin(p_i(\vec n)) x_i/\hbar)\\
\chi&=&\sqrt{\frac{2^3}{V}}\,\chi_0^+(\vec p(\vec n))\, e^{-i E_B(\vec n) t/\hbar}
\prod_{i=1}^3 \sin(p_i(\vec n)) x_i/\hbar)
\eea
in which $p_i(\vec n)=n_i\pi\hbar/L_i$ while  $\varphi_0^+(\vec p(\vec n))$ and $\chi_0^+(\vec p(\vec n))$ are
given by (\ref{phi0}) and (\ref{chi0}) respectively, with $E_p$ replaced
by $E_B(\vec n)$. Once again the charge current $\vec J$ is zero.
%
%
%
\section{Comparison between the energy spectra for spin-0 and spin-1/2 relativistic particles in a box}
%
%
\subsection{One-dimensional box}
%
%
The energy solutions of the Klein-Gordon and Dirac equation for particles in a one-dimensional box of size $L$ are given,
in terms of the allowed wave numbers $k_n$, by (see \cite{afg})
\bea
\label{KG_BC_box1D}\hbox{Klein-Gordon}\ &&E_B(n)= \sqrt{\hbar^2 c^2 k_n^2 + m^2c^4 }\  ,\  \ k_n=n\pi/L\ ,\\
\label{Dirac_BC_box1D}\hbox{Dirac}\ &&E_B(n)= \sqrt{\hbar^2 c^2 k_n^2 + m^2c^4 }\  ,\ \ \tan(k_n L)=-\frac{\hbar k_n}{mc} \ .
\eea
In the case of the Dirac equation, $k_n$ denotes the $n$th solution (excluding the trivial one $k=0$) of the
transcendental equation. It is convenient to write these quantities in terms of dimensionless quantities and consider the
scaled kinetic energy ${\cal T}=T/(mc^2)
=E/(mc^2)-1$
instead of the total energy. Defining $x_n=k_n\lambda_C$ and $L_C=L/\lambda_C$, where $\lambda_C=\hbar/(mc)$ is the Compton wavelength
of the relativistic particle of mass $m$, one gets
\bea
\hbox{Klein-Gordon}\ &&{\cal T}^{KG}_B(n)= \sqrt{x_n^2 + 1}-1\ ,\quad x_n=n\pi/L_C\ ,\\
\label{Dirac_box1D}{\rm Dirac}\ &&{\cal T}^D_B(n)= \sqrt{x_n^2 + 1}-1\ ,\quad\ \tan(L_C x_n)=-x_n \ .
\eea
One sees that the main difference between Klein-Gordon and Dirac spectra is the value of the quantity $x_n$.
Writing the transcendental equation for $x_n$ in (\ref{Dirac_box1D}) as
\beq
\tan(y_n)=-\frac{y_n}{L_C}\,, \qquad y_n=x_n L_C\ ,
\eeq
it is clear that, for lengths of the box much bigger than the Compton wavelength, i.e., $L_C\gg 1$, one has $y_n\sim n\pi$ and thus $x_n\sim n\pi/L_C$. Since, on the other hand, $x_n$, at least for values
of $n$ not too high, would be small in this case, we would be in the non-relativistic limit. Therefore, one can state that the effect of spin on the
energy of a relativistic particle in a box is itself coming from relativity, or the covariance of the Dirac equation, since this is how the Dirac plane wave spinor is constructed, and thus the origin of the transcendental equation above. This is similar to what happens to a spin-1/2 particle in a spherical box, for which the spin-orbit coupling disappears for big (non-relativistic) boxes \cite{afo}.

The difference between  the two equations (\ref{KG_BC_box1D}) and (\ref{Dirac_BC_box1D}) can be traced back to 
the boundary conditions obeyed by the wave function for a particle-in-a-box problem in Klein-Gordon and Dirac equations, 
respectively. In the case of Klein-Gordon equation, being a second-order differential equation in the coordinates, one 
requires, as in the Schr\"odinger equation, that the wave function vanishes at the box borders because of the continuity of the wave function. 
However, this cannot be done in the Dirac equation, as mentioned before and discussed in refs. \cite{afg} and \cite{adv}. 
Indeed, being a first-order differential equation in the coordinates, the former requirement cannot be made, and it is replaced 
by a boundary condition which is equivalent to demand that the current flux $\vec j\cdot\hat n$ vanishes at the
 box borders ($\hat n$ is the outward unit vector at the borders). Of course, this happens in the Klein-Gordon equation 
as well -- the current vanishes at the borders -- but in the Dirac equation the current $\vec j$ is fundamentally 
different from the Klein-Gordon one because of the spinor structure of the wave function. 
Therefore, this structure, related to the spin 1/2 nature of the Dirac particles, affects the energy spectrum 
for a spin-1/2 relativistic particle in a box problem. It turns out that both boundary condition are equivalent in the non-relativistic limit, because in this limit one recovers the non-relativistic solutions of the same problem (see \cite{afg, adv}). This spin effect in relativistic problems does not happen necessarily with other interactions, namely with the interaction with a pure vector electromagnetic field (except for the Stern-Gerlach term) or for certain combination of scalar and vector potentials \cite{prc_75_047303}.

For the electron, $\lambda_C\sim 3.86\times 10^{-13}\ {\rm m}=3.86\times 10^{-3}$ {\AA},
 a relatively small value at atomic scale, but one could envisage an electron confined in smaller boxes
than an atom-sized box, given a sufficient high electric field. For spin-0 mesons like the charged pions $\pi^\pm$,
 one has $\lambda_C\sim 1.41\ {\rm fm}=1.41\times 10^{-5}$ {\AA}.

In the following figure we plot the values of ${\cal T}^{KG}_B(n)$ and  ${\cal T}^{D}_B(n)$ for $n=1,\ldots,4$,  in logarithmic scale,
for several values of $L_C$.
The non-relativistic value ${\cal T}^{nr}_B(n)=x_n^2/2$ is presented for the highest value of $L_C=300$, which is about $1.15$ {\AA} for an electron.
\begin{figure}[!ht]
\begin{center}
\includegraphics[width=12cm]{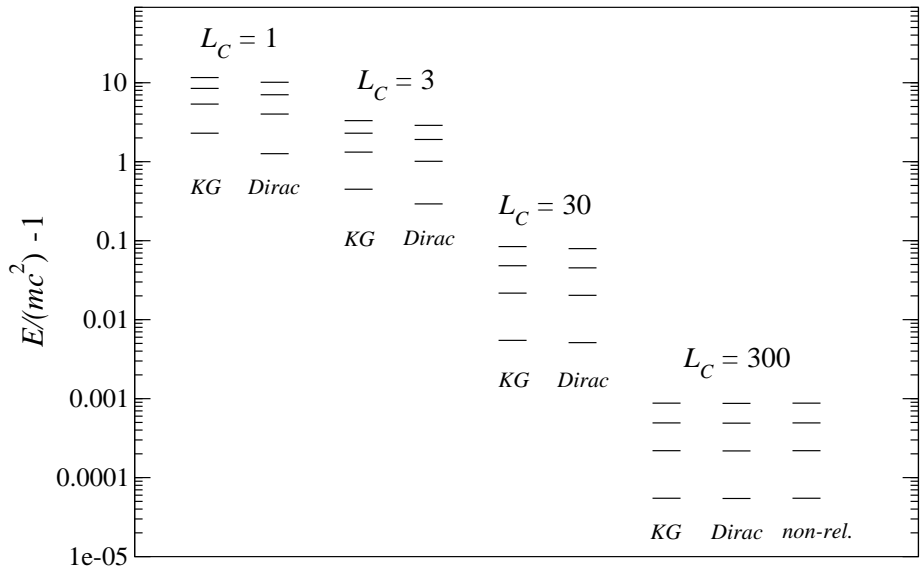}
\end{center}
\caption{Scaled kinetic energy spectrum of the first four solutions of Klein-Gordon and Dirac equations
for four values of the ratio $L_C$ for a one-dimensional box.}
\label{Fig1}
\end{figure}
%
%
%
\subsection{Three-dimensional box}
%
%
%
%
For the three-dimensional infinite well we have now for the Klein-Gordon and Dirac energies \cite{adv}
\bea
\hbox{Klein-Gordon}\ && E_B(\vec n)= \sqrt{\hbar^2 c^2 |\vec k(\vec n)|^2 + m^2c^4 }\ ,\ k_i(\vec n)=\frac{n_i\pi}{L_i}\ ,\ i=1,2,3\ ,\\
{\rm Dirac}\ && E_B(\vec n)= \sqrt{\hbar^2 c^2 |\vec k(\vec n)|^2 + m^2c^4 }\ ,\nonumber\\
&&\hspace*{0.0cm}\tan(k_i(\vec n) L_i)=-\frac{2(E_B(\vec n)+mc^2)\hbar k_i(\vec n)\,c}
{\hbar^2c^{2} k_i^2(\vec n)-(E_B(\vec n)+mc^2)^2}\ ,\ i=1,2,3 \ .
\eea
As in the previous section, $\vec n$ denotes the set of quantum numbers $(n_1,n_2,n_3)$, all non-zero positive integers.
In the case of the Dirac equation, they are used to
label the different solutions of the three coupled transcendental equations, as discussed in \cite{adv}.
Considering now a cubic box, i.e., $L_1=L_2=L_3=L$, and, as in the one-dimensional case, using the scaled kinetic energy and defining
$x_i(\vec n)=k_i(\vec n)\lambda_C$, $L_C=L/\lambda_C$, we get
\bea
\hspace*{-.7cm}\hbox{Klein-Gordon}&&\hspace*{-.3cm}{\cal T}^{KG}_B(\vec n)= \sqrt{|\vec x(\vec n)|^2 + 1}-1\,,
\ x_i(\vec n)=\frac{n_i\pi}{L_C}\,,\  i=1,2,3\ ,\\
\hbox{Dirac}&&\hspace*{-.3cm}{\cal T}^D_B(\vec n)= \sqrt{|\vec x(\vec n)|^2 + 1}-1\ ,\nonumber \\
&&\tan(x_i(\vec n) L_C)=\frac{2({\cal T}^D_B(\vec n)+2)x_i(\vec n)}
{x_i^2(\vec n)-({\cal T}^D_B(\vec n)+2)^2}\ ,\ i=1,2,3 \ .
\eea
Again, defining $y_i(\vec n)=x_i(\vec n)L_C$ one has for the transcendental equations
\beq
\tan(y_i(\vec n))=L_C\frac{2({\cal T}^D_B(\vec n)+2)y_i(\vec n)}
{y_i^2(\vec n)-L_C^2({\cal T}^D_B(\vec n)+2)^2} \ .
\eeq
When $L_C\gg 1$, one has $\tan(y_i(\vec n))\sim 0$ and thus $x_i\sim n_i\pi/L_C$, and thus, as before, we get the Klein-Gordon
(and in this case, non-relativistic) result. It is interesting to remark that, if one $x_i(\vec n)$ is much bigger than the others, then
$x_i(\vec n)\sim|\vec x(\vec n)|$ and the respective equation becomes
\bea
\tan(x_i(\vec n) L_C)&=&\frac{2({\cal T}^D_B(\vec n)+2)x_i(\vec n)}
{x_i^2(\vec n)-({\cal T}^D_B(\vec n)+2)^2}\nonumber\\
&=& \frac{2({\cal T}^D_B(\vec n)+2)x_i(\vec n)}
{x_i^2(\vec n)-|\vec x(\vec n)|^2-2({\cal T}^D_B(\vec n)+2)}\sim -x_i(\vec n)
\eea
exactly as the one-dimensional case.

For the 3-dimensional box, there is an energy degeneracy regarding the levels with quantum numbers $(n_1,n_2,n_3)$: any permutation
of the $n_i$'s gives rise to the same energy, as it is evident from the Klein-Gordon energy expression (\ref{KG_box3D}) for a
cubic box and was already noted in \cite{adv} for a Dirac particle in a cubic box as well. Therefore, one has 3 types of degeneracies:
1-fold for $n_1=n_2=n_3$; 3-fold when two $n_i$'s are the same but the third is different;
$3!=6\,$-fold degeneracy when the $n_i$'s are all different.

The following figure is similar to Fig.~\ref{Fig1} for the three-dimensional box. In this case the first four distinct energy levels correspond to $\vec n$ being equal to
$(1,1,1)$, $(1,1,2)$, $(1,2,2)$ and $(1,1,3)$ in increasing energy order. Because of the degeneracies mentioned above, one is actually depicting 10 energy levels. Actually, in the Dirac case, one still has an additional 2-fold degeneracy for each level, corresponding to the two independent spin polarizations.

\begin{figure}[!ht]
\begin{center}
\includegraphics[width=12cm]{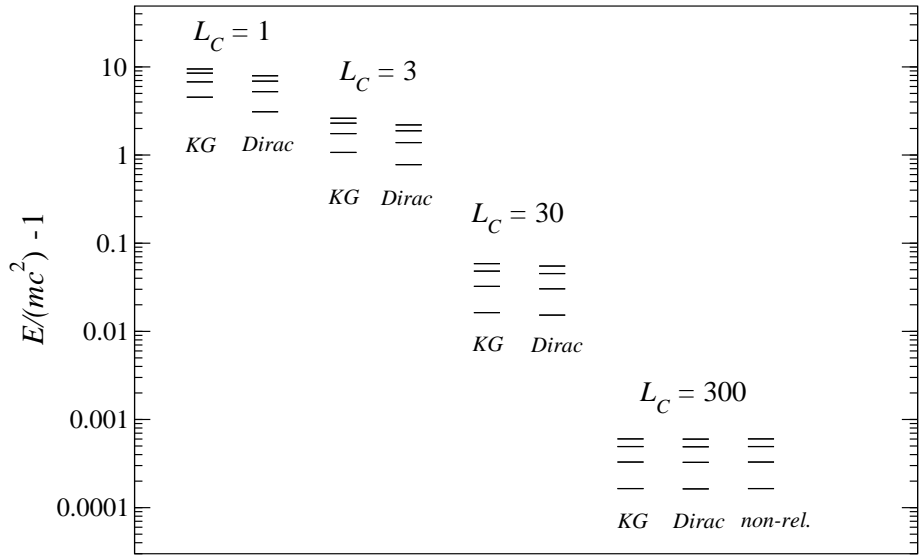}
\end{center}
\caption{Scaled kinetic energy spectrum of the first four solutions of Klein-Gordon and Dirac equations
for four values of the ratio $L_C$ for a three-dimensional box.}
\label{Fig2}
\end{figure}

Several aspects are common to the Klein-Gordon and Dirac on
three-dimensional box spectra: the energy levels for spin-1/2 are in general lower than the ones for spin-0 particles and,
for larger boxes, measured in units of the Compton wavelength, they approach the non-relativistic values, as could be expected, since larger boxes mean smaller momenta for the particle confined.
In the three-dimensional case, there are more levels up to a certain energy and they are closer to each other than in the one-dimensional case.
This can be understood by the fact that with increasing $n$, the number of combinations of $(n_1,n_2,n_3)$ such that
$n_1^2+n_2^2+n_3^2\leq n^2$ increases more than linearly with $n$, actually as $n^3$ for sufficiently high $n$.

As discussed in \cite{adv}, the energy levels computed for a three-dimensional box allows us to compare the density of states of a
relativistic fermion gas to the density of a non-relativistic fermion gas. Since
the separation between $k_i$ values is smaller than $\pi/L_i$ for volumes $V\sim \lambda_C^3$, the relativistic density is higher
than the non-relativistic one. This can also be seen in Fig.~\ref{Fig2}. For a spin-0 gas, in spite of the fact the separation between allowed
$k_i$ value is identical with the non-relativistic case, one must take into account the Bose-Einstein statistics, by which one has,
at zero temperature, as many states as one wishes in the same energy state.

\section{Conclusions}
%
%
In this paper, we have solved the problem of a relativistic spin-0 particle in a one- and three-dimensional
box using the Klein-Gordon equation in the Feshbach-Villars formalism. By using position-dependent mass for the
infinite square well potential we were able to avoid the Klein paradox problem and thus retain a one-particle description
of spin-0 particles.
We compare these solutions with those previously found for the Dirac equation with the same potentials.
We find that the spin has indeed an effect on the energy spectra for relativistic spin-1/2 particles as compared with
relativistic spin-0 particles with the same mass, decreasing the energy of the former compared in the latter.
As expected, both kinds of particles tend to the same non-relativistic spectrum, if the size of the box is sufficiently
large in units of Compton wavelength.

%

\vskip1cm
\noindent\textbf{Acknowledgments}
\vskip.1cm
\par\noindent
This work was supported in part by the Natural
Sciences and Engineering Research Council of Canada.



\end{document}